\documentclass[twocolumn,showpacs,amssymb,preprintnumbers,prl,floatfix]{revtex4}
\usepackage{rotating,epsfig,dcolumn}
\usepackage{graphicx,bm}

\begin{document}
\title{A Shell Model Description of the Decay Out of the  Super-Deformed Band of $^{36}$Ar}
\author{E. Caurier $^a$, F. Nowacki $^a$ and
   A.~Poves $^b$}
\affiliation{
(a) IReS, B\^at27, IN2P3-CNRS/Universit\'e Louis
Pasteur BP 28, F-67037 Strasbourg Cedex 2, France\\
(b) Departamento de Fisica Te\'orica, C-XI. Universidad Aut\'onoma
de Madrid, E-28049, Madrid, Spain}
\date{\today}
\begin{abstract}
  Large scale shell model calculations in the valence space spanned by two major oscillator shells
  ($sd$ and $pf$) describe simultaneously the super-deformed excited
  band of $^{36}$Ar and its spherical ground state. 
  We explain the appearance
  of this super-deformed band at low excitation energy  as a
  consequence of the very large quadrupole correlation energy of the
  configurations with  many
  particles and many holes (np-nh) relative to the normal filling of the
  spherical mean field orbits (0p-0h).
   We study the mechanism of mixing between the different
  configurations, to understand why the super-deformed band survives
  and how it finally decays into the low-lying spherical states via
  the indirect mixing of the 0p-0h and 4p-4h configurations.

\end{abstract}
\maketitle

 Excited deformed bands in spherical nuclei provide  a spectacular
 example of coexistence of very different structures at the same
 energy scale, that is a rather peculiar aspect of the dynamics of the atomic
 nucleus. Several cases are known since long, 
 for instance the four particle four holes and eight particles eight
 holes states in $^{16}$O, starting at
 6.05~MeV and 16.75~MeV of excitation energy \cite{O16a,O16b}.
 However, it is only recently that
 similar bands, deformed and even super-deformed, have been
 discovered in other medium-light nuclei such as
 $^{56}$Ni~\cite{ni56exp}, 
 $^{36}$Ar~\cite{ar36exp} and  $^{40}$Ca~\cite{ca40exp} and explored
 up to high spin.
 These experiments have been possible thanks to the advent of large
 arrays of $\gamma$ detectors, like Gammasphere or Euroball. 
 One characteristic feature of these bands is that they belong to
 rather well defined spherical shell model configurations; for
 instance, the deformed excited band in $^{56}$Ni can be associated with
 the configuration (1f$_{7/2}$)$^{12}$ (2p$_{3/2}$, 1f$_{5/2}$,
 2p$_{1/2}$)$^{4}$ while the (super)deformed bands in  $^{36}$Ar and  $^{40}$Ca
 are based in structures $(sd)^{16}$ $(pf)^{4}$ and  $(sd)^{16}$
 $(pf)^{8}$ respectively. The location
 of the np-nh states in  $^{40}$Ca was studied in the Hartree-Fock
 approximation with blocked particles and Skyrme forces in ref.~\cite{zamick}.
 While many approaches are
 available for the microscopic description of these bands (Cranked
 Nilsson-Strutinsky \cite{ar36exp}, Hartree-Fock plus BCS with
 configuration mixing \cite{bender}, Angular Momentum Projected Generator Coordinate Method \cite{rayner},
  Projected Shell
 Model \cite{sun}, Cluster models \cite{sakuda} etc.) the
 interacting shell model is, when affordable, the prime choice.
  The  mean field description of N=Z nuclei, has 
 problems related to the proper treatment of the
 proton-neutron pairing in its isovector and isoscalar channels. On
 the shell model side, the problems come from the size of the valence
 spaces needed to accommodate the np-nh configurations. 


  In this letter we focus  in the $^{36}$Ar case, where a rotational band has been experimentally found
  starting 4.3~MeV above the spherical ground state. It is
  generated by  an intrinsic super-deformed state {\it i.e.} bearing an axis ratio 2:1. In
  previous works \cite{ar36exp,svens2}, we have shown that the
  experimental super-deformed (SD) band can be associated to the promotion
  of four particles across the Fermi level, from the $sd$ to the $pf$ shell. Indeed, this is
  a very crude description, because the physical states  contain
  components belonging to other np-nh  configurations.
  This mixing allows for 
  transitions connecting the superdeformed band and the low-lying $sd$
  states. However, the mixing must be gentle enough so as not to
  jeopardize  the very existence of  the band. This is the
  theoretical challenge that we affront in this letter: to explain why
  the super-deformed band appears at such low excitation energy  and
  how can, at the same time, mix with other configurations  and  preserve its identity.


 Let's recall the results of the calculations at fixed 4p-4h configuration.
 In the description of the SD band of $^{36}$Ar,
 the natural valence space comprises the $sd$ and the $pf$ shells. 
 However, the inclusion of the 1d$_{5/2}$ orbit 
 produces a huge increase in the size of the
 basis and massive center of mass effects, forcing us
 to exclude it from the valence space.
  The quadrupole coherence of the solutions will be reduced
 by this truncation. We have quantified the effect and it is
 moderate as we shall show below.
 Therefore, our valence space consists of  the
 orbits  2s$_{1/2}$, 1d$_{3/2}$, 1f$_{7/2}$, 2p$_{3/2}$, 1f$_{5/2}$,  and
 2p$_{1/2}$. The basis dimensions are O(10$^7$). The starting effective interaction is the same used in
 ref.~\cite{ar36exp} dubbed {\sf SDPF.SM} and described in detail in \cite{rmp}.
 Blocking the orbit 1d$_{5/2}$ minimizes the center of mass contamination of
 the solutions. Nevertheless, to be safer, we have added to the interaction the
 center of mass hamiltonian with a
 small coefficient $\lambda_{cm}$=0.5 to reduce even more unwanted
 mixings (see
 \cite{dean:1999} for an updated discussion of the center of mass issues). The
 differences between the
 results with and without the addition of the
 center of mass hamiltonian are minor.

\begin{figure}
  \resizebox{0.45\textwidth}{!}{%
  \includegraphics{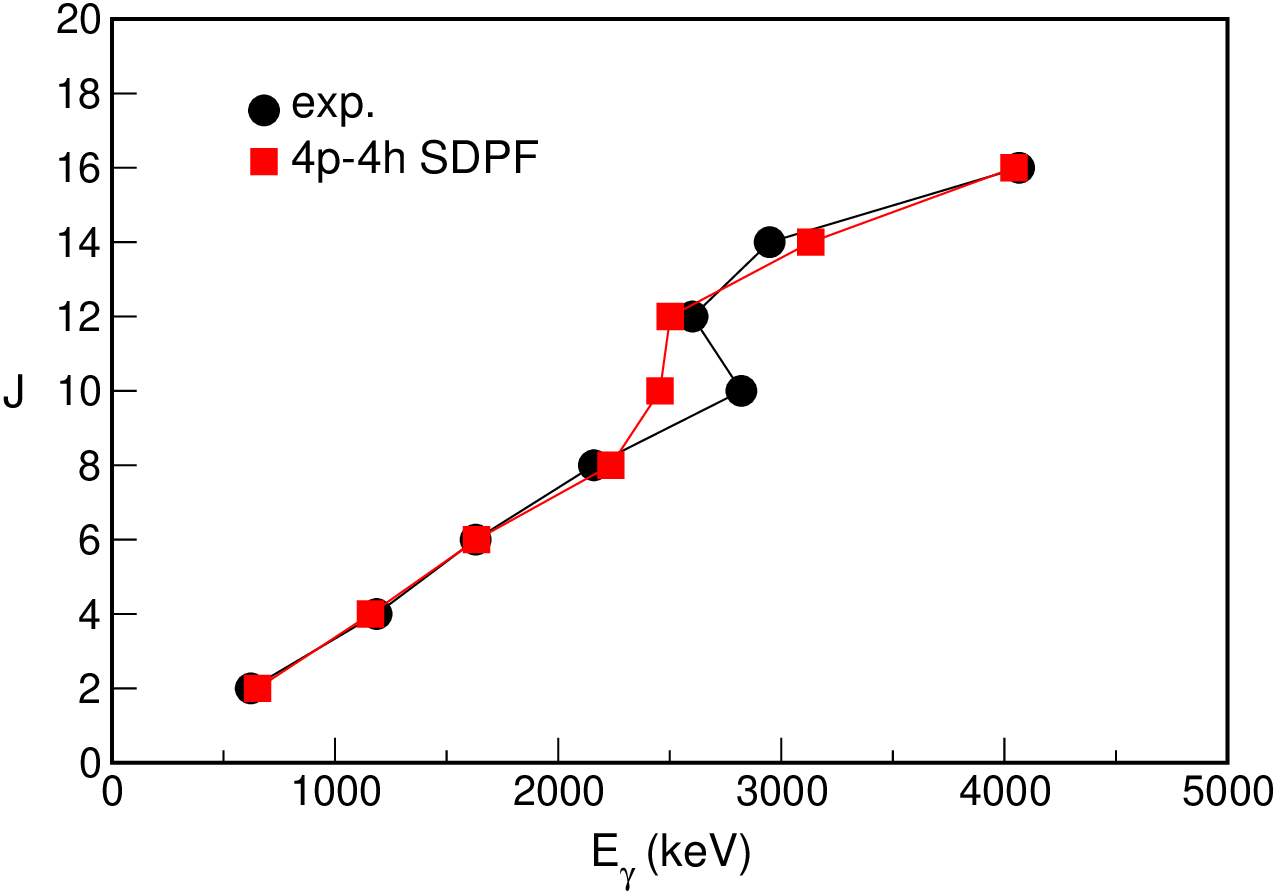}}
 \caption{The superdeformed band in $^{36}$Ar; E$_{\gamma}$'s, exp. vs. 4p-4h calculation}
 \label{fig:1}       
 \end{figure}

\begin{figure}
  \resizebox{0.45\textwidth}{!}{%
  \includegraphics{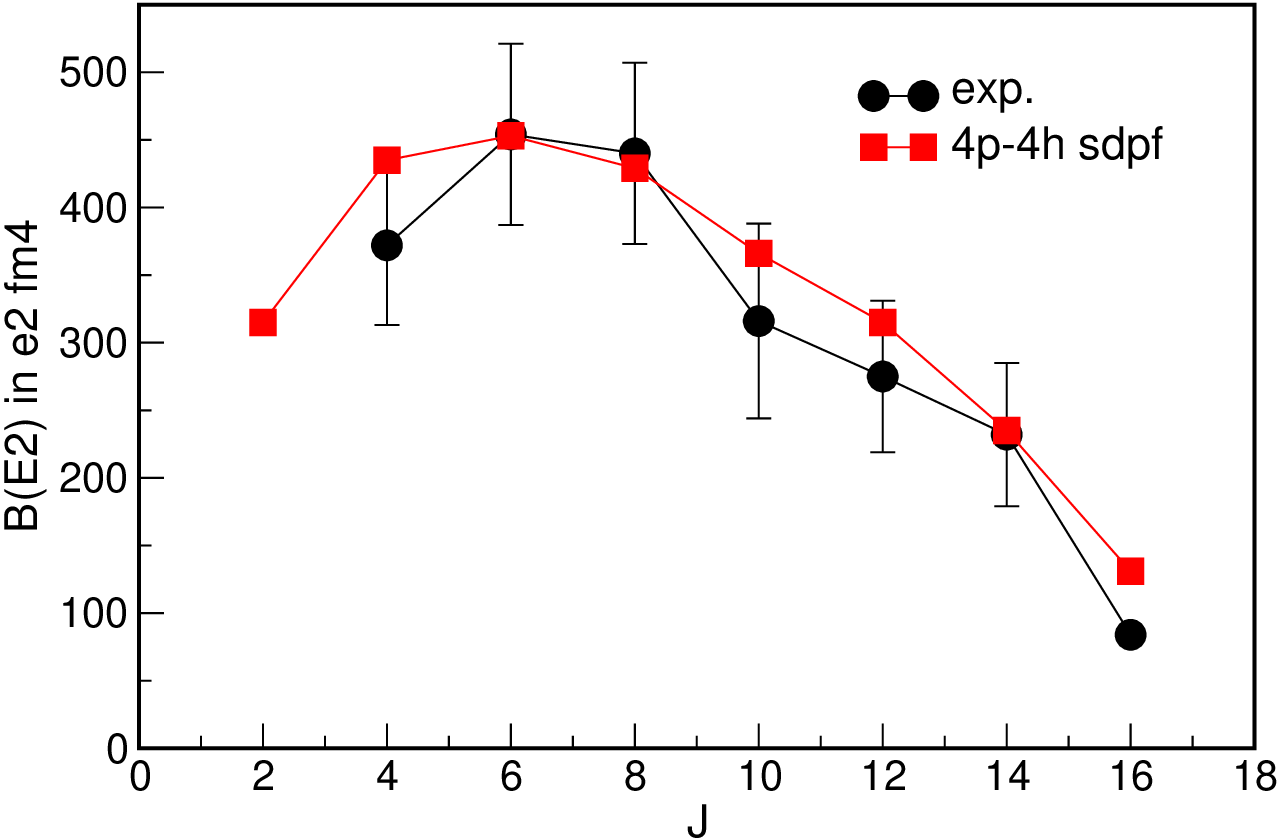}} 
 \caption{The superdeformed band in $^{36}$Ar; B(E2)'s, exp. vs. 4p-4h
   calculation. The J=16 point is only a lower bound}
 \label{fig:2}       
 \end{figure}

 In Figure~\ref{fig:1} the calculated transition energies are 
 compared with
 the experimental results in a backbending plot. The agreement is
 really  remarkably good, except at J=12 where the data show a clear backbending
 while the calculation produces a much smoother upbending pattern. In
 the experimental data there is a close-by second 10$^+$ state,
 therefore, the discrepancy may be due to the lack of mixing in the
 calculation. If we move now to 
 Fig.~\ref{fig:2} where we gather the experimental~\cite{svens2} and calculated B(E2)'s
 (we use standard effective charges $\delta q_{\pi}$=$\delta
 q_{\nu}$=0.5 and b=1.94~fm), we find again an  
 astonishing accord\footnote{ Actually it can be seen to be
better than it was in \cite{svens2}. The reason is
that there, the default value of the oscillator size parameter
in  the  codes; b=1.01 A$^{1/6}$ fm, which is a 5\% smaller than the
correct one for these light nuclei, was inadvertently used.}.
The value of the intrinsic quadrupole moment corresponds
roughly to a deformation $\beta$=0.5, so that it is justified to speak of a 
super-deformed band. The origin of such a strong coherence can be attributed to the
dominance of Elliott's SU(3) structures in  the valence space.
 The almost perfect agreement achieved by the 4p-4h  description, is
 at the same time a blessing  and a challenge. Surprisingly, the incorporation
 of all the other degrees of freedom must result in
 essentially no modification at all of the results. And this is not an
 easy task \footnote{As the Queen  says to the protagonist of ``Alice through the looking
 glass'': ``Now, here, you see, it takes all
 the {\sl running} you can do to keep in the same place $\ldots$''}.


 The interaction {\sf SDPF.SM} was originally devised for calculations in fixed N$\hbar
 \omega$ spaces. In a mixed calculation, some changes must be enforced.
 In the first place, we must retire from the pairing matrix elements
 of the $pf$-shell the implicit effect of the cross shell excitations
 that now are taken explicitly into account. The amount of this
 reduction can be estimated in second order perturbation theory to be
 roughly the square of the off-diagonal matrix element (2~MeV)
 divided by the gap between the 0p-0h and the 2p-2h states (8~MeV). We
 model this effect by reducing all the $pf$ shell T=1 pairing
 matrix elements a 40\%. Moreover, the closure of the 1d5/2 orbit
 affects differently to the correlation energies of the np-nh
 configurations and we must correct the calculations for this
 limitation. To quantify the effect, we have performed calculations in
 the complete $sd-pf$ space for the 2p-2h and 4p-4h configurations (in
 this last case for high-J members of the band). The results show that
 the 2p-2h and 4p-4h configurations
 gain 1.5~MeV and 5.0~MeV more than the 0p-0h configuration when the
 1d5/2 orbit is opened. In order to simulate this effect, we have
 added to the $pf$-shell  monopole interaction a term  $\delta
 \epsilon \; n_{pf} \; + \; 1/2 \; \delta \overline{V} \;
 n_{pf}^{(2)}$, with $\delta \epsilon$~=~--~0.5~MeV
 and  $\delta \overline{V}$~=~--~0.5~MeV. This prescription
 overestimates the correlation energy of the 8p-8h states, but it
 has no relevant influence in the states we are aiming to.
 Concerning the B(E2)'s, in the case of the 4p-4h band, the opening of
 the 1d5/2 orbit produces a 12\% increase of their values.
 

 \begin{figure}
  \resizebox{0.45\textwidth}{!}{%
  \includegraphics{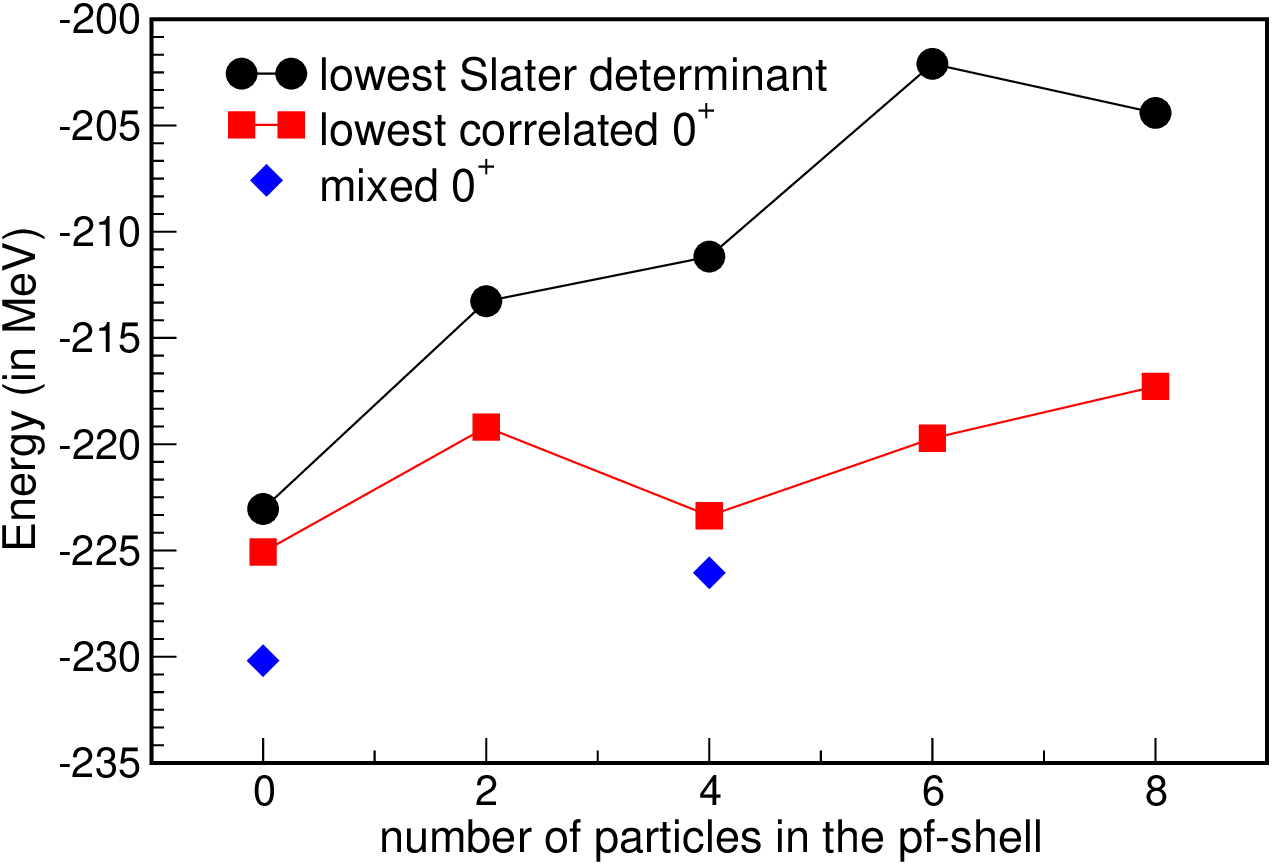}}
 \caption{Relative position of lowest states of 
   the np-nh configurations. Diamonds
   represent the energies of the ground state and the superdeformed
   band-head after mixing}
 \label{fig:3}       
 \end{figure}

  Once the interaction has been adapted to the valence space, we
  proceed to make calculations --using the shell model code {\sf
  ANTOINE}  \cite{rmp,antoine}-- in spaces of fixed np-nh
  character. Naively one would locate these states at an excitation
  energy equal to n times the quasi-particle gap between the $sd$ and
  the $pf$ orbits ($\sim$7~MeV). In Fig.\ref{fig:3} this
  corresponds to a straight line with slope 7~MeV. This estimation
  misses two crucial  ingredients; first, the presence of quadratic terms in the
  monopole hamiltonian that can lower substantially the uncorrelated energy of these
  configurations, and second, the gain in energy of the maximally correlated states within these
  configurations. To measure the size of the first contribution, we have
  computed the energies of the lowest Slater Determinant for each np-nh
  space. We have plotted them in Fig.~\ref{fig:3}. Notice that the
  behavior is not linear; in particular the 4p-4h
  configuration does not lay  28~MeV above the 0p-0h, but just  12~MeV (see
  \cite{zuker.mono} for a fully worked example of this
  mechanism). The next step is to diagonalize the full interaction in the
  np-nh spaces to incorporate the correlations at fixed configuration.  
  The results are the squares in   Fig.~\ref{fig:3} and the correlation
  energy can be defined as the difference  between circles and
  squares in the figure. The key point is that these correlation
  energies are very large. In
  the 4p-4h space, they are large enough to produce a super-deformed 
  rotor whose correlation energy exceeds that of the
  spherical ground state by 15~MeV, a huge effect that represents 5\% of the total binding energy.
  This suffices to bring the 4p-4h
  band-head close to  degeneracy with the 0p-0h state. It is important
  to notice that the combined effect of the monopole hamiltonian and
  the correlation energy  favors clearly the 4p-4h super-deformed
  band which becomes the lowest configuration above the 0p-0h ground
  state. Both the 2p-2h and the 6p-6h band-heads are higher in
  energy. This has important consequences for the mechanisms of mixing.


 \begin{figure}
  \resizebox{0.45\textwidth}{!}{%
  \includegraphics{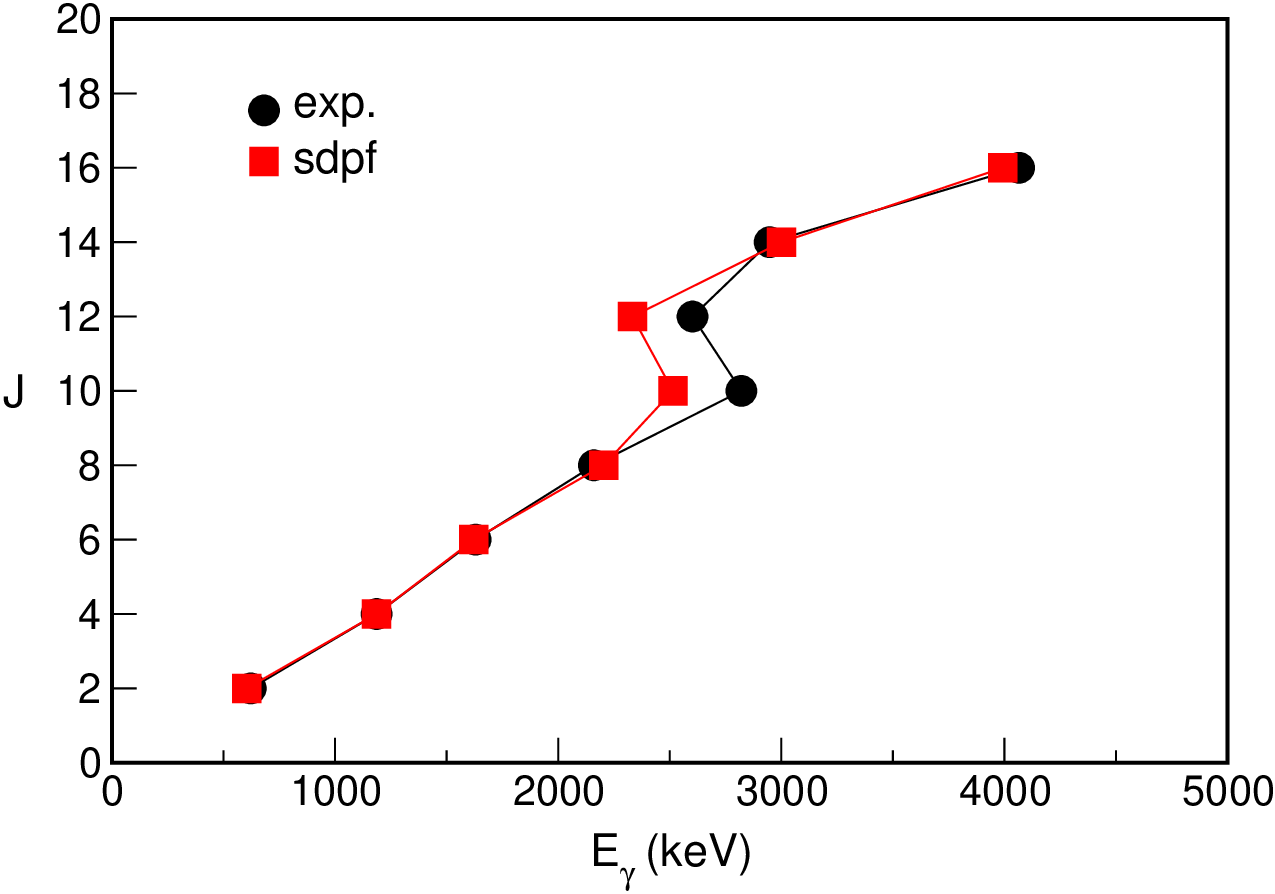}}
 \caption{The superdeformed band in $^{36}$Ar; E$_{\gamma}$'s,
   comparison with the results of the mixed calculation}
 \label{fig:4}       
 \end{figure}

 The final step consist in the diagonalization  of the interaction in the full
 valence space.
  As a consequence of the mixing with the 2p-2h states,
 the 0p-0h dominated (70\%)
 ground state gains 5~MeV,  while the 4p-4h dominated (70\%) band-head
 of the super-deformed band gains 2.5~MeV (see Fig.~\ref{fig:3}), to finish close to its
 experimental location (4329~keV exp $vs.$ 4319~keV th.). The relative placement of the configurations
 at fixed np-nh and the intrinsic  structure of the np-nh states produce
 the following  very interesting effect: 
 The 4p-4h super-deformed states mix mostly with the 6p-6h ones. A
 larger mixing with the 2p-2h states would have prompted an earlier
 and more intense decay-out of the super-deformed band. The 0p-0h and
 2p-2h states are of single particle (or spherical nature) and the
 cross shell pairing interaction --$i.e.$ the scattering of a nuclear Cooper pair
 from the $sd$ to the $pf$ shell-- mixes them  very efficiently. The
 lower 4p-4h and 6p-6h states are both super-deformed and pairing mixes
 them too. What turns out to be  severely hindered is the mixing of the spherical
 2p-2h states with the super-deformed 4p-4h states, as semi-classical
 arguments suggest. In addition, the
 states of the super-deformed band cannot decay to those of the 6p-6h
 band because they lay at higher energies. The decay out of the
 super-deformed band
 must then proceed via its small 0p-0h and 2p-2h components.

   \begin{figure}
  \resizebox{0.45\textwidth}{!}{%
  \includegraphics{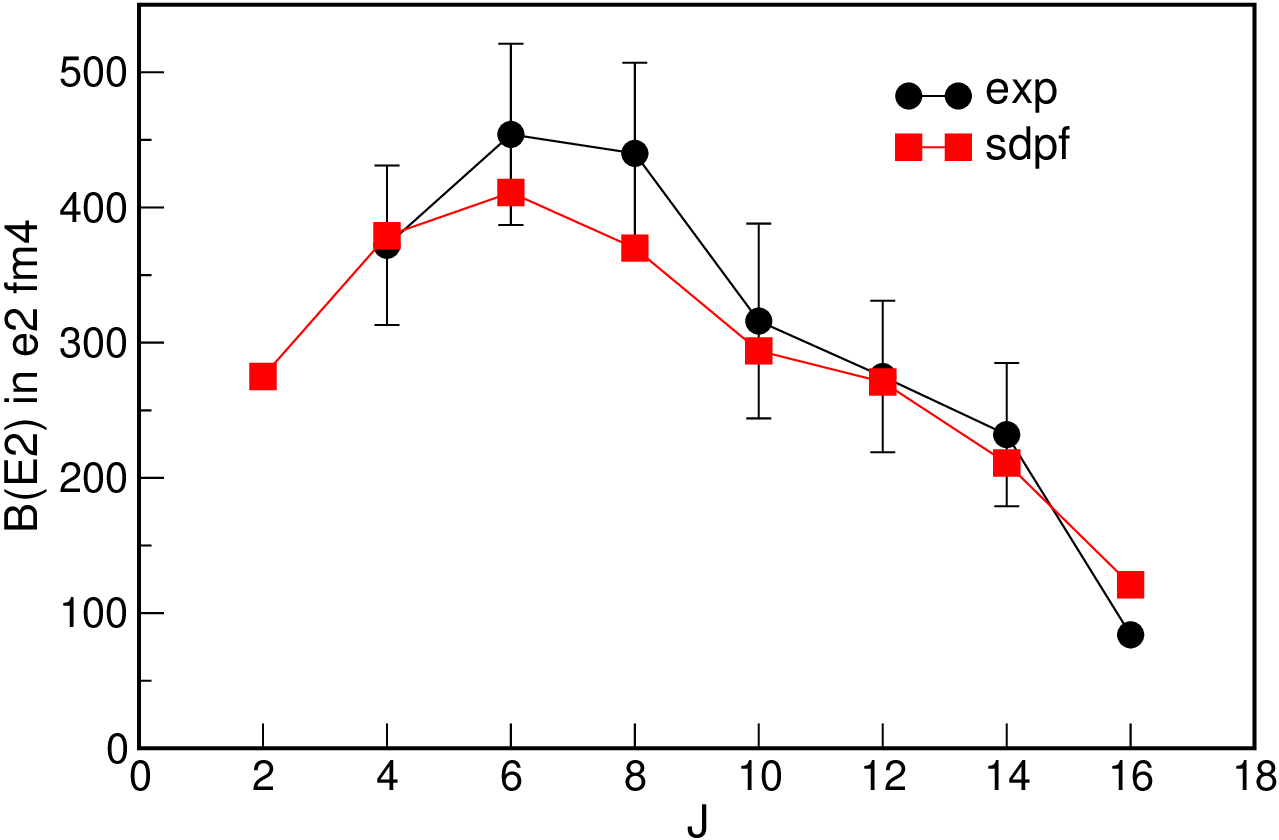}}
 \caption{The superdeformed band in $^{36}$Ar; B(E2)'s, exp. vs. mixed
 results.}
 \label{fig:5}       
 \end{figure}

 In Fig.~\ref{fig:4} we compare the $\gamma$ energies along the SD
 band with the experiment. Even if it seemed difficult to improve upon
 the agreement of the 4p-4h calculation, the mixed calculation
 succeeds. Now the backbending is better reproduced while the moment of inertia is
 still correct. For the first, the interaction of the SD 10$^+$ with
 an isolated 2p-2h 10$^+$ does the job. For the second, the
 off-diagonal pairing, active in the mixed calculation, compensates the
 pairing reduction of the effective interaction (as it should). In the
 range of sensible values of the pairing interaction there is surely
 room to absorb the hypothetical modifications of the moment of
 inertia of the SD band due to the opening of the 1d5/2 orbit. 

  In Fig.~\ref{fig:5} we compare the experimental and theoretical
  B(E2)'s through the SD band. The accord is excellent and would become
  perfect should we increase the calculated values by 12\%, the quantity
  that we have obtained as  the effect of the opening of the 1d5/2
  in our fragmentary  4p-4h calculations. The reconstruction of the
  4p-4h results is due to the fact that the 4p-4h SD states mix with
  6p-6h states that are in turn almost as deformed as them. A larger
  mixing with the (spherical) 2p-2h states would have produced an
  unwanted reduction of the B(E2)'s.


\begin{table}[h]
\caption{{\label{tab:ar36}}
 Out-band transitions in the SD band of  $^{36}$Ar (B(E2)'s in
 e$^2$fm$^4$ and energies in keV)}
\begin{ruledtabular}
\begin{tabular}{ccccc}
 & \multicolumn{2}{c} {E$_{\gamma}$} & \multicolumn{2}{c}{ B(E2)}  \\
  &  experiment &  theory & experiment  & theory \\   
\hline
 2$^+_{SD}$ $\rightarrow$ 0$^+_1$  & 4950   & 4846 & 4.6(23)  & 4.0 \\
 4$^+_{SD}$ $\rightarrow$ 2$^+_1$  & 4166   & 3917 & 2.5(4)   & 1.2 \\
 4$^+_{SD}$ $\rightarrow$ 2$^+_2$  & 1697   &  946 & 19.2(30) & 18.4 \\
 6$^+_{SD}$ $\rightarrow$ 4$^+_1$  & 3552   & 2787 & 5.3(8)   & 0.25 \\
10$^+_1$ $\rightarrow$ 8$^+_{SD}$  & 1975   & 1192 & 43.6(74) & 13.1 \\
12$^+_{SD}$ $\rightarrow$ 10$^+_1$ & 3448   & 3655 & 15.0(30) & 1.5 \\
\end{tabular} 
\end{ruledtabular}
\end{table}

 In table \ref{tab:ar36} we have gathered the experimental results
 obtained in  \cite{svens2} for the transitions connecting states of
 the SD band with other states of 0p-0h or 2p-2h character.  The
 decay-out of the band occurs through the transitions  
 2$^+_{SD}$ $\rightarrow$ 0$^+_1$,
 4$^+_{SD}$ $\rightarrow$ 2$^+_1$ and 6$^+_{SD}$ $\rightarrow$ 4$^+_1$,
 that share two characteristics; a very small transition strength
 and a large energy release. The calculation reproduces fairly well
 these properties, thus giving a microscopic explanation of the
 decay-out. At the bottom of the SD band, the energies of the in-band
 emitted $\gamma$'s become  small, and the phase space enhancement of the
 transitions to the low lying spherical states compensates  the
 smallness of the transition strengths. In the upper part of the band,
 the presence of a 2p-2h 10$^+$ yrast state. produces a weak out-band
 excursion that is also well given by the calculation.

  The mean field calculation of ref.~\cite{bender} is  possibly the 
  closest to ours. Albeit it is  limited to the low spin regime
  J$\le$6, it  incorporates the mixing of different projected mean
  field solutions using the Generator Coordinate Method. As we have
  mentioned before, the standard mean field calculations do not treat properly the proton
  neutron pairing. In the cranking approximation this results in a too large
  moment of inertia. Indeed, as shown for the $^{48}$Cr case
  in  ref.~\cite{cr48}, the correct treatment of the neutron
  proton pairing halves the moment of inertia.
  Paradoxically, the moments of inertia of
  the SD bands calculated in \cite{bender} are too small. According to
  the authors this is a drawback of their method, related to absence of
  $\Delta$K=0 admixtures, as discussed in \cite{rayner0}.
  This spurious reduction of the moment of inertia overcompensates 
  the effect of their  bad treatment of the proton neutron pairing.
  The  cancellation of this two --unphysical-- but opposite contributions can
  eventually produce an 
  unwanted  ``correct'' moment of inertia. As pointed out by the authors of 
  ref.~\cite{rayner}, this is actually the case in their AMP-CGM
  calculation using the Gogny force.
    The in-band 
  B(E2)'s of the SD band are  clearly overestimated in \cite{bender}
  while for the transitions out of the band their predictions are
  acceptable but less accurate than ours.

  To complete the description we have calculated a side band of
  negative parity with  a  3$^-$ band-head at 4178~keV,
  E$_{\gamma}$(5$^-$ $\rightarrow$  3$^-$)=~993~keV and
  E$_{\gamma}$(6$^-$ $\rightarrow$  5$^-$)=~2183~keV also reported in
  \cite{svens2}. In order to describe the negative parity states we find that it is
  compulsory to open the 1d5/2 orbit. As a full calculation is out of
  reach, we have computed the negative parity states in the space of
  the 1p-1h and 3p-3h configurations, referring them to the ground
  state energy calculated in the space of the 0p-0h and 2p-2h
  configurations. The results are  reasonable  with the  3$^-$ at
  5040~keV , E$_{\gamma}$(5$^-$ $\rightarrow$  3$^-$)=~751~keV and
  E$_{\gamma}$(6$^-$ $\rightarrow$  5$^-$)=~2010~keV. Notice that
  $^{36}$Ar exhibits at around 4~MeV a multiplet of nearly degenerate
  states of 0p-0h, 1p-1h and 4p-4h nature, that the calculations are
  able to explain simultaneously.


  In conclusion, we have shown that ``state of the art'' shell model
  calculations in the valence space spanned by two major oscillator
  shells can provide an unified description of many coexisting
  nuclear structures; spherical states of single particle nature,
  negative parity states of octupole character and a super-deformed
  band built on  4p-4h excitations.
  In particular, we have shown why the super-deformed band  appears at
  such a low excitation energy and how the indirect mixing of very
  different states governs the decay-out of the super-deformed band.

This work is partly supported by the IN2P3(France) CICyT(Spain)
collaboration agreements. AP's work is supported by  MCyT (Spain),
 grant BFM2003-1153.

\end{document}